\documentclass[preprint,aps,prd,amsmath,amssymb,nofootinbib]{revtex4}
\usepackage{graphicx}
\usepackage{color}
\usepackage[latin1]{inputenc}
\usepackage{ulem}
\newcommand{\be}{\begin{eqnarray}}
\newcommand{\beq}{\begin{equation}}
\newcommand{\eeq}{\end{equation}}
\newcommand{\ee}{\end{eqnarray}}
\newcommand{\nn}{~\nonumber \\}
\newcommand{\p}{{\cal P}\exp}
\newcommand{\texp}{{\cal T}\exp}
\newcommand{\ssh}{\gamma\cdot}
\newcommand{\im}{{\rm Im}}
\newcommand{\bmp}{\noindent\begin{minipage}{16cm}}
\newcommand{\emp}{\end{minipage}\vskip 7mm} 
\newcommand{\unsplit}{\check}
\newcommand{\system}{\hat}
\newcommand{\kernel}{\bar}
\newcommand{\sigmaf}{\sigma\hspace{-1mm}:\hspace{-1mm}F}
\usepackage{graphicx}
\usepackage{dcolumn}
\usepackage{bm}
\usepackage{amsmath}
\usepackage{amsfonts}
\usepackage{bbm}
\usepackage{subfigure}

\newcommand{\drawsquare}[2]{\hbox{%
\rule{#2pt}{#1pt}\hskip-#2pt
\rule{#1pt}{#2pt}\hskip-#1pt
\rule[#1pt]{#1pt}{#2pt}}\rule[#1pt]{#2pt}{#2pt}\hskip-#2pt
\rule{#2pt}{#1pt}}
\newcommand{\Yfund}{\raisebox{-.5pt}{\drawsquare{6.5}{0.4}}}
\newcommand{\Yasymm}{\raisebox{-3.5pt}{\drawsquare{6.5}{0.4}}\hskip-6.9pt%
                     \raisebox{3pt}{\drawsquare{6.5}{0.4}}%
                    }
\newcommand{\Ysymm}{\Yfund\hskip-0.4pt%
                    \Yfund}
\def\symm{\Ysymm}
\def\bsymm{\overline{\Ysymm}}
\def\drawbox#1#2{\hrule height#2pt
        \hbox{\vrule width#2pt height#1pt \kern#1pt
              \vrule width#2pt}
              \hrule height#2pt}
\def\Fund#1#2{\vcenter{\vbox{\drawbox{#1}{#2}}}}
\def\Asym#1#2{\vcenter{\vbox{\drawbox{#1}{#2}
              \kern-#2pt 
              \drawbox{#1}{#2}}}}
\def\sym#1#2{\vcenter{\hbox{ \drawbox{#1}{#2} \drawbox{#1}{#2} }}}
\def\fund{\Fund{6.4}{0.3}}
\def\asymm{\Asym{6.4}{0.3}}
\def\bfund{\overline{\fund}}
\def\basymm{\overline{\asymm}}

\newcommand{\Dsl}{\not\!\!D}
\newcommand{\rem}{{\bf !!!:}}
\def\s{{\,\rm s}}
\def\g{{\,\rm g}}
\def\eV{\,{\rm eV}}
\def\keV{\,{\rm keV}}
\def\MeV{\,{\rm MeV}}
\def\GeV{\,{\rm GeV}}
\def\TeV{\,{\rm TeV}}
\def\sv{\left<\sigma v\right>}
\def\({\left(}
\def\){\right)}
\def\cm{{\,\rm cm}}
\def\K{{\,\rm K}}

\begin{document}

\title{Composite dark matter from a model with composite Higgs boson}
\author{Maxim Yu. {\sc Khlopov}}\email{khlopov@apc.univ-paris7.fr}
\author{Chris {\sc Kouvaris}}\email{kouvaris@nbi.dk}
 \affiliation{$^*$Center for Cosmoparticle physics ``Cosmion",
125047, Moscow, Russia\\
Moscow Engineering Physics Institute, 115409 Moscow, Russia, and \\
APC laboratory 10, rue Alice Domon et Léonie Duquet 75205 Paris
Cedex 13, France, \\
 $^{\dagger}$The Niels Bohr Institute, Blegdamsvej 17, DK-2100 Copenhagen,
Denmark}

\begin{abstract}
In a previous paper~\cite{Khlopov:2007ic}, we showed how the minimal
walking technicolor model (WTC) can provide a composite dark matter
candidate, by forming bound states between a $-2$ electrically
charged techniparticle and a $^4He^{++}$. We studied the properties
of these \emph{techni-O-helium} $tOHe$ ``atoms'', which behave as
warmer dark matter rather than cold. In this paper we extend our
work on several different aspects. We study the possibility of a
mixed scenario where both $tOHe$ and bound states between $+2$ and
$-2$ electrically charged techniparticles coexist in the dark matter
density. We argue that these newly proposed bound states solely made
of techniparticles, although they behave as Weakly Interacting
Massive Particles (WIMPs), due to their large elastic cross section
with nuclei, can only account for a small percentage of the dark
matter density. Therefore we conclude that within the minimal WTC,
composite dark matter should be mostly composed of $tOHe$. Moreover
in this paper, we put cosmological bounds in the masses of the
techniparticles, if they compose the dark matter density. Finally we
propose within this setup, a possible explanation of the discrepancy
between the DAMA/NaI and DAMA/LIBRA findings and the negative
results of CDMS and other direct dark matter searches that imply
nuclear recoil measurement, which should accompany ionization.

\end{abstract}


\maketitle
\section{Introduction}

Walking technicolor theories (WTC) have regained a lot of interest
recently. This is because they can naturally break the electroweak
symmetry without violating experimental constraints by the
electroweak precision measurements. Several technicolor theories
that have techniquarks transforming under higher representations
of the gauge group, require a small number of colors and flavors
in order to become
quasi-conformal~\cite{Sannino:2004qp,Hong:2004td,Dietrich:2005jn,Dietrich:2006cm}.
Because of this property, the Higgs particle can be composed of
two techniquarks and be able to couple even to the heavier
Standard Model particles like the top quark. On the other hand,
the fact that these theories become conformal only for a small
number of colors and flavors, differentiates them from the old
baroque technicolor models that are excluded by the electroweak
precision measurements. In addition, the possibility of
unification of the couplings makes the walking technicolor
theories legitimate candidates for the Large Hadronic Collider
(LHC)~\cite{Gudnason:2006mk}.

Among the walking technicolor theories, special interest has been
drawn to the minimal case. This particular model contains only two
techniquarks that transform under the adjoint representation of
the $SU(2)$ technicolor group and a new lepton family in order to
cancel the Witten global anomaly. This minimal model has been
investigated thoroughly in~\cite{Gudnason:2006ug,Foadi:2007ue}. A holographic description
of the theory was presented in~\cite{Dietrich:2008ni}, where several predictions regarding
the mass spectrum were made. Lattice methods have also been used recently for
the study of gauge theories with fermions that transform under higher dimensional representations~\cite{Catterall:2007yx,DelDebbio:2008wb,DelDebbio:2008zf}.
Although simple in nature, this minimal walking technicolor model
can provide several possibilities for dark matter. In particular,
the theory can admit as dark matter particles
technibaryons~\cite{Gudnason:2006yj}, bound states between a
neutral techniquark and a technigluon~\cite{Kouvaris:2007iq},
heavy leptons of the fourth lepton
family~\cite{Kainulainen:2006wq}, or bound states between a
 $-2$ electrically charged techniparticle and a $He^{++}$~\cite{Khlopov:2007ic}.
In the latter case, WTC offers a new exciting realization of a
composite dark matter scenario, which was earlier considered in
different aspects in the model of teraparticles
\cite{Glashow:2005jy,Fargion:2005xz}, in the AC model
\cite{Fargion:2005ep,Khlopov:2006uv}, based on the approach of an
almost commutative geometry \cite{Connes:1994yd,Stephan:2005uj},
and in the model of 4th generation
\cite{Khlopov:2005ew,Belotsky:2006fd,Belotsky:2006pp,4Q}, assuming
existence of stable heavy $U$ quark \cite{Belotsky:2005ui}.

In all these recent models (see review in \cite{Khlopov:2006dk,Khlopov:2007zza,Khlopov:2008rp,Khlopov:2008rq}),
the predicted stable charged particles form neutral atom-like states, composing the dark matter of the modern Universe
and escaping experimental discovery. It offers
new solutions for the physical nature of the cosmological dark
matter. The main problem for these solutions is to suppress the abundance of positively charged
species bound with ordinary electrons, which behave as anomalous isotopes of hydrogen or helium.
This problem remains unresolved, if the model predicts stable particles with charge $-1$,
as it is the case for tera-electrons \cite{Glashow:2005jy,Fargion:2005xz}.

The possibility of stable doubly charged particles $A^{--}$ and
$C^{++}$, revealed in the AC model, offered a candidate for dark
matter in the form of elusive (AC)-atoms. In the charge symmetric case, when primordial concentrations of $A^{--}$ and $C^{++}$ are equal, their binding in the expanding Universe is not complete due to freezing out and a significant fraction of free   relic $C^{++}$, which is not bound in (AC)-atoms, is left in the Universe and represents a potential danger of anomalous helium overproduction. The suppression of this fraction in terrestrial matter involves a new long range interaction between A and C, making them to recombine in (AC)-atoms inside dense matter bodies~\cite{Fargion:2005ep,Khlopov:2006uv}.

In the asymmetric case, corresponding to excess of $-2$ charge
species, as it was assumed for $(\bar U \bar U \bar U)$ in the
model of stable $U$-quark of a 4th generation, their positively
charged partners annihilate effectively in the early Universe. The
dark matter is in the form of nuclear interacting O-helium -
atom-like bound states of $-2$ charged particles and primordial
helium, formed as soon as $He$ is produced in Big Bang
Nucleosynthesis (BBN). Such an asymmetric case was realized in
\cite{Khlopov:2007ic} in the framework of WTC, where it was
possible to find a relationship between the excess of negatively
charged anti-techni-baryons and/or technileptons and the baryon
asymmetry of the Universe.

The minimal walking technicolor model we use
is the same as in our previous paper~\cite{Khlopov:2007ic} (and
references therein). It contains two techniquarks that transform
under the adjoint representation of an $SU(2)$ gauge group, i.e.
up $U$ and down $D$, with electric charges $1$ and $0$
respectively. There is also a new fourth family of leptons $\nu'$
and $\zeta$ with charges $-1$ and $-2$ respectively. This
hypercharge assignment is not unique, however it is consistent,
since it makes the theory gauge anomaly free.
It was already noticed in \cite{Khlopov:2007ic} that since two types of stable doubly charged particles
(technibaryon $(UU)^{++}$ and technilepton $\zeta^{--}$) can exist, the excess of positively charged $(UU)^{++}$
together with the excess of negatively charged $\zeta^{--}$ is also possible, giving rise to
atom-like $[(UU)\zeta]$ WIMP species.

Here we analyze the ability of WTC to provide this WIMP solution
for composite dark matter.
It is evident that the predicted abundance and cosmological role
of $[(UU)\zeta]$ are determined by the relation between the
excess of its constituents $(UU)^{++}$ and $\zeta^{--}$. Their
excess can be different, although the case where the excess of
$(UU)^{++}$ is larger, leads to the unresolved problem of
anomalous helium overproduction.

One can find a similar problem in the case where the excess of
$(UU)^{++}$ is equal to the excess of $\zeta^{--}$. In full
analogy with the cosmology of the AC model
\cite{Fargion:2005ep,Khlopov:2006uv}, most of $(UU)^{++}$ and
$\zeta^{--}$ are bound in this case in $[(UU)\zeta]$ ``atoms", but
the remaining fraction of unbound $(UU)^{++}$ is still up to ten
orders of magnitude larger than the experimental upper limits on
anomalous helium in terrestrial matter \cite{exp3}. Since the
minimal WTC can not offer new long range interactions between
$(UU)$ and $\zeta$, ordinary atoms of anomalous helium $[(UU)ee]$
and nuclear interacting techni-O-helium $[He^{++}\zeta^{--}] $,
having different mobilities in matter, inevitably fractionate. It
prevents their recombination in $[(UU)\zeta]$, which might reduce
the concentration of anomalous helium in terrestrial matter below
experimental upper limits.

Therefore to solve the problem of anomalous helium in the
framework of minimal WTC, we are left with the only option to have
the excess of negatively charged $\zeta^{--}$ larger than the
excess of $(UU)^{++}$. This provides complete binding of
$(UU)^{++}$ in $[(UU)^{++}\zeta^{--}]$, while the residual
excessive $\zeta^{--}$ bind with helium in techni-O-helium. This
solution can be effective even if the excess of $\zeta^{--}$
exceeds the excess of $(UU)^{++}$ by relative amount of $\sim
10^{-8}$. Therefore it seems that the WIMPs $[(UU)\zeta]$ can be
the dominant dark matter component, making the nuclear interacting
techni-O-helium dynamically negligible, as it was the case for the
AC model \cite{Fargion:2005ep,Khlopov:2006uv}.

However, we'll show here that unlike the neutral $(AC)$ atoms,
having zero electroweak charge, the weak charge of $[(UU)\zeta]$ is
non-zero and its interaction with nuclei, mediated by ordinary
$Z$-boson, should lead to an observable effect in the CDMS
experiment  \cite{Akerib:2005kh,Ahmed:2008eu}, unless the
contribution of $[(UU)\zeta]$ to the total dark matter density is
restricted to be a few percent.

An interesting feature of the considered scenario is that in a wide interval of masses
 of $(UU)$ and $\zeta$, the generation of
excess corresponding to the saturation of the observed dark matter
by techniparticles, predicts a fixed negative value for the ratio
of lepton number $L$ over the baryon number $B$. This ratio is
constant for masses below few TeV and then rapidly grows by
absolute value for larger masses and exceeds $10^{8}$, when they
approach 10 TeV. A large negative value of $L/B$ corresponds to
strong lepton asymmetry and to the excess of antineutrino in the
period of BBN, which leads to a corresponding growth of primordial
$He$ abundance. This argument provides an upper limit on masses of
techniparticles.

The paper is organized as follows. After a brief description of
the general chronological framework for the considered techniparticle
Universe (Section \ref{Chronology}), we study the relation
between baryon asymmetry and techniparticle excess, fixing the
value of $L/B$ ratio (Section \ref{Excess}).
In Section \ref{Detection}, we deduce an upper limit on possible
contributions of $[(UU)\zeta]$ WIMPs in the total dark matter
density, which follow from the most recent severe constraints of the
CDMS experiment \cite{Ahmed:2008eu}. We also speculate on the
possibility to explain the positive results of DAMA/NaI (see for
review \cite{Bernabei:2003za}) and DAMA/Libra \cite{Bernabei:2008yi}
experiments by ionization effects of inelastic processes, induced by
techni-O-helium in the matter. We consider the main results of the
present work in Section \ref{Discussion}.

\section{\label{Chronology} Chronological framework of the techniparticle Universe}

It is well known that strong technicolor interactions provide
strong exponential suppression of frozen antitechnibaryons, if a
technibaryon excess is generated. Since the technilepton
interaction is much weaker, even in the presence of a technilepton
excess, the freeze out concentration of positively charged
anti-techni-leptons $\bar \zeta^{++}$ can be significant. However,
our previous detailed analysis \cite{Khlopov:2007ic} has shown
that in the period after BBN, all the remaining $\bar \zeta^{++}$
can be effectively eliminated by techni-O-helium catalysis. This
catalysis, taking place after all the free $\zeta^{--}$ bind with
helium, formed in BBN, provides also an effective binding of all
the remaining free $(UU)$ in $[(UU)\zeta]$ WIMPs. The constraint
on the contribution of $[(UU)\zeta]$ WIMPs to the total dark
matter density, which we deduce in Section \ref{Detection} from
the results of CDMS search for WIMPs, makes the dynamical
evolution of the considered techniparticle Universe virtually
coinciding with the picture of techni-O-helium Universe, studied
in \cite{Khlopov:2007ic}.

On the above reasons we can avoid a detailed analysis of all the
stages of cosmological evolution of techniparticles and give only
a brief sketch of this evolution, which serves as a framework for
our further discussion of several specific problems.

The thermal history of techniparticles starts with the generation
of baryon (and/or lepton) asymmetry in the very early Universe.
The mechanism of such generation is not specified in the minimal
WTC, but owing to sphaleron processes, this asymmetry is
redistributed in the equilibrium, giving rise to the excess of
technilepton $L'$, technibaryon $TB$, baryon and lepton numbers.
After the freeze out of the sphaleron processes, these numbers are
conserved separately and the excess of technibaryons and
technileptons is fixed. It results to the excess of the lightest
stable technibaryons $UU^{++}$ and technileptons $\zeta^{--}$.
Both species behave as charged leptons in particle physics
experiments and their absence in accelerator searches puts a lower
bound on their mass, about 100 GeV.  For numerical estimations
below we introduce the notation $m_{\zeta} = 100 S_2{\GeV} $ for
the mass of $\zeta^{--}$, $m_{UU} = 100 B_2{\GeV} $ for the mass
of $UU^{++}$ and $\mu=m_{UU}m_{\zeta}/(m_{UU}+m_{\zeta})=100
R_2{\GeV} $ for the reduced mass of the $UU$ and $\zeta$ system. With
the use of this notation, the chronology of techniparticle
evolution after the generation of technibaryon and technilepton
asymmetry, looks as follows:

1) In the period $10^{-10}S_2^{-2}\s \le t \le 6
\cdot10^{-8}S_2^{-2}\s$ at $m_{\zeta} \ge T \ge T_f=m_{\zeta}/31
\approx 3 S_2 \GeV$, $\zeta$-lepton pairs $\zeta \bar \zeta$
annihilate and freeze out. For large $m_{\zeta}$, the abundance of
frozen out $\zeta$-lepton pairs is not suppressed in spite of a
$\zeta$-lepton excess. A similar period with the exchange of $S_2$
by $B_2$ can be mentioned for the freeze out of $UU$ and $\bar U
\bar U$ pairs. Due to the strong technicolor interaction, the
freeze out abundance of $\bar U \bar U$ is exponentially small for
all the reasonable masses of technibaryons.

Even at the largest possible values of $S_2$ and $B_2$, the freeze out
temperature $T_f$ for techniparticles does not exceed
substantially the freeze out temperature for sphaleron
processes. Because of this, the process of freezing out the
technipartcles can not strongly influence the conditions under
which techniparticle excess is generated.

2)In the period $6 \cdot 10^{-4}R_2^{-2}\s \le t \le 5.4
R_2^{-2}10^{-1}\s$ at $I_{U\zeta} \approx 40 R_2 \MeV \ge T \ge
I_{U\zeta}/30 $ negatively charged technileptons $\zeta^{--}$ can
recombine with positively charged technibaryons $UU^{++}$ in
atom-like atoms $[(UU)\zeta]$ with potential energy $I_{U\zeta}=
Z_{UU}^2 Z_{\zeta}^2\alpha^2 \mu/2 \approx 40 R_2 \MeV $
($Z_{UU}=2$, $Z_{\zeta}=-2$). This process is frozen out at $T
\approx I_{U\zeta}/30,$ when the inverse reaction of $[(UU)\zeta]$
photodestruction is not effective to prevent recombination of $UU$
and $\zeta$. Together with neutral $[(UU)\zeta]$ atoms, free charged
$\zeta^{--}$ and $UU^{++}$ are also left, being the dominant form of
techniparticle matter at large $R_2$.

3)At $ t \sim 2.4 \cdot 10^{-3}S_2^{-2}\s$  and the temperature $T
\sim I_{\zeta} = 20 S_2 \MeV,$ corresponding to the binding energy
$I_{\zeta} = Z_{\zeta}^4 \alpha^2 m_{\zeta}/4 \approx 20 S_2 \MeV$
($Z_{\zeta}=-2$) $\zeta$-positronium ``atoms" $(\zeta^{--} \bar
\zeta^{++})$ are formed, in which $\bar \zeta^{++}$ annihilate. At
large $m_{\zeta}$ this annihilation is not at all effective to
reduce the $\zeta \bar \zeta$ pairs abundance. These pairs are
eliminated in the course of the successive evolution of
techniparticles.

4)In the period $100\s \le t \le 300\s$  at $100 \keV\ge T \ge
T_o=I_{o}/27 = 8 \alpha^2 m_{He}/27 \approx 60 \keV$, $^4He$ has already been formed in the
BBN and virtually all free $\zeta^{--}$ are trapped by $^4He$ in
techni-O-helium ``atoms" $(^4He^{++}\zeta^{--})$. Being formed,
techni-O-helium catalyzes the binding of free $UU$ with its
constituent $\zeta^{--}$ in $[(UU)\zeta]$ atoms and of free $\bar
\zeta^{++}$  into $\zeta$-positronium and completes the annihilation of
all the primordial antitechnileptons. At large $m_{\zeta}$, in spite
of a significant fraction of free $\bar \zeta^{++}$, the effects of
$(\zeta^{--} \bar \zeta^{++})$ annihilation catalyzed by
techni-O-helium, do not cause any contradiction with observations.

Techni-O-helium, being an $\alpha$-particle with screened electric
charge, can catalyze nuclear transformations, which can influence
primordial light element abundance and cause primordial heavy
element formation. These effects need a special detailed and
complicated study. The arguments of \cite{Khlopov:2007ic} indicate
that this model does not lead to immediate contradictions with the
observational data.

After having been formed, the weakly interacting neutral $UU\zeta$ ``atoms"
immediately decouple from the plasma, being close to typical cold dark matter
particles by spectrum of their density fluctuations.

Due to nuclear interactions of its helium constituent with nuclei
in the cosmic plasma, the techni-O-helium gas is in thermal
equilibrium with plasma and radiation on the Radiation Dominance
(RD) stage, while the energy and momentum transfer from the plasma
is effective. The radiation pressure acting on the plasma is then
effectively transferred to density fluctuations of the
techni-O-helium gas and transforms them in acoustic waves at
scales up to the size of the horizon. However, as it was first
noticed in \cite{Khlopov:2005ew}, this transfer to heavy
nuclear-interacting species becomes ineffective before the end of
the RD stage and such species decouple from plasma and radiation.
Consequently, nothing prevents the development of gravitational
instability in the gas of these species. This argument was shown
in \cite{Khlopov:2007ic} to be completely applicable to the case
of techni-O-helium.

5) At temperature $T < T_{od} \approx 45 S^{2/3}_2\eV$,
estimated in \cite{Khlopov:2007ic}, the energy and
momentum transfer from baryons to techni-O-helium is not effective
because $n_B \sv (m_p/m_o) t < 1$, where $m_o$ is the mass of the
$tOHe$ atom, $m_p$ is the mass of the proton, and $S_2=\frac{m_o}{100 \GeV}$. Here \beq \sigma
\approx \sigma_{o} \sim \pi R_{o}^2 \approx
10^{-25}\cm^2\label{sigOHe}, \eeq where $R_o$ is the size of
techni-O-helium, $n_B$ is the baryon number density, and $v = \sqrt{2T/m_p}$ is the
baryon thermal velocity. The techni-O-helium gas decouples from
the plasma and plays the role of dark matter, which starts to
dominate in the Universe at $T_{RM}=1 \eV$.

6) Therefore in the period after $t \sim 10^{12}\s$  at $T \le
T_{RM} \approx 1 \eV$, the
 techniparticle dominance starts. Due to the CDMS constraints \cite{Ahmed:2008eu}
 (see Section \ref{Detection}), the allowed fraction of $UU\zeta$ is small and the techni-O-helium ``atoms"
play the main dynamical role in
the development of gravitational instability, triggering the
large scale structure formation.
The composite nature of techni-O-helium determines the specifics of the
corresponding dark matter scenario.

The total mass of the $tOHe$ gas with density $\rho_d =
\frac{T_{RM}}{T_{od}} \rho_{tot}$ within the cosmological horizon
$l_h=t$ is
$$M=\frac{4 \pi}{3} \rho_d t^3.$$ In the period of decoupling $T = T_{od}$, this mass  depends
strongly on the techniparticle mass $S_2$ and is given in \cite{Khlopov:2007ic}\beq
M_{od} = \frac{T_{RM}}{T_{od}} m_{Pl} (\frac{m_{Pl}}{T_{od}})^2
\approx 2 \cdot 10^{46} S^{-8/3}_2 \g = 10^{13} S^{-2}_2
M_{\odot}, \label{MEPm} \eeq where $M_{\odot}$ is the solar mass, and $m_{Pl}$ is the Planck mass.
The techni-O-helium is formed only at $T_o$ and its total mass
within the cosmological horizon in the period of its creation is
$M_{o}=M_{od}(T_{od}/T_{o})^3 = 10^{37} \g$.

On the RD stage before decoupling, the Jeans length $\lambda_J$ of
the $tOHe$ gas was restricted from below by the propagation of
sound waves in plasma with a relativistic equation of state
$p=\epsilon/3$, being of the order of the cosmological horizon and
equal to $\lambda_J = l_h/\sqrt{3} = t/\sqrt{3}.$ After decoupling
at $T = T_{od}$, it falls down to $\lambda_J \sim v_o t,$ where
$v_o = \sqrt{2T_{od}/m_o}.$ Though after decoupling the Jeans mass
in the $tOHe$ gas correspondingly falls down
$$M_J \sim v_o^3 M_{od}\sim 3 \cdot 10^{-14}M_{od},$$ one should
expect a strong suppression of fluctuations on scales $M<M_o$, as
well as adiabatic damping of sound waves in the RD plasma for scales
$M_o<M<M_{od}$. It can provide some suppression of small scale
structure in the considered model for all reasonable masses of
techniparticles. The significance of this suppression and its effect
on the structure formation needs a special study in detailed
numerical simulations. In any case, it can not be as strong as the
free streaming suppression in ordinary Warm Dark Matter
scenarios, but one can expect that qualitatively we deal with a Warmer
Than Cold Dark Matter model. Being decoupled from baryonic matter,
the $tOHe$ gas does not follow the formation of baryonic
astrophysical objects (stars, planets, molecular clouds...) and
forms dark matter halos of galaxies.

Based on this general framework, we analyze the generation of
techniparticle excess and the results of direct dark matter
searches, which fix the parameters of the considered scenario and
lead to robust predictions.

\section{\label{Excess} Calculation of techniparticle excess}

 In this section we
calculate the relic densities of the particles of interest. Our
derivation is similar as in~\cite{Gudnason:2006yj,Khlopov:2007ic}.
We assume the existence of a
baryon-antibaryon asymmetry created sometime after inflation.
Equilibrium between ordinary and techni-particles, supported by WTC,
provides a definite relation between this asymmetry and
technibaryon-antitechnibaryon asymmetry. In addition, we assume that
the conditions of thermal equilibrium via weak interactions hold down to the temperature where
the sphalerons freeze out. Under these conditions, the ratio of the
technibaryon number $TB$ over the baryon number $B$ is given as \beq
\frac{TB}{B}=-\sigma_{UU}\left(\frac{L'}{B}\frac{1}{3\sigma_{\zeta}}+1+\frac{L}{3B}\right)
\label{tbb},\eeq where $L$ and $L'$ are the lepton number and the
technilepton number respectively. The parameters $\sigma_{UU}$ and
$\sigma_{\zeta}$ are statistical factors that depend on the mass of
the particle in question and the freeze out temperature of the
sphalerons and are defined in~\cite{Khlopov:2007ic}. We should also
mention that the above equation holds under the condition of a
second order phase transition for the electroweak symmetry. The
results for a first order phase transition has been shown not to
change the conclusions significantly~\cite{Gudnason:2006yj}.

In~\cite{Khlopov:2007ic}, we examined the possibility that
negatively $-2$ charged particles with substantial abundance can
bind with $^4He$ forming electrically neutral atoms that can play
the role of strongly interacting massive particles (SIMP) as part of
warm dark matter. In particular, we investigated the possibility of
$He\bar{U}\bar{U}$ and $He\zeta$ atoms. In this paper, we continue
our investigation by looking at the possibility that $UU^{++}$ can
bind with $\zeta^{--}$ forming again a neutral atom that can
contribute to the dark matter density. As we are going to argue,
such a bound state differs qualitatively from the $He\zeta$ and $He\bar{U}\bar{U}$ states,
since it behaves more as a WIMP, rather than a SIMP. As it has been
argued previously~\cite{Gudnason:2006yj,Khlopov:2007ic}, both $UU$
and $\zeta$ carry technibaryon $TB$ and technilepton number $L'$
respectively. If no interactions violate $TB$ and $L'$, and if $UU$
and $\zeta$ are the lightest technibaryon and technilepton
respectively, these particles are absolutely stable. Here, we
investigate the possibility that there is a substantial relic
density for $UU$ and $\zeta$, with the density of $\zeta$ being
higher than that of $UU$. In this case, the overwhelming majority of
$UU$ will be captured by $\zeta$ forming neutral bound states
$UU\zeta$, while the remaining $\zeta$ will be captured by
$^4He^{++}$, forming neutral $He\zeta$. In such a case, the
contribution to the dark matter density $\Omega_d$ is \beq
\frac{\Omega_d}{\Omega_B}=\frac{\Omega_{TB}}{\Omega_{B}}+\frac{\Omega_{L'}}{\Omega_B}=
\frac{3}{2}\frac{TB}{B}\frac{m_{d}}{m_p} +
\left(\frac{L'}{B}-\frac{3}{2}\frac{TB}{B} \right )\frac{m_{\zeta}}{m_p},\label{darko} \eeq where $m_d$ is the mass of
the bound state of $UU\zeta$, $m_p$ is the mass of the proton, and
$m_{\zeta}$ is the mass of the bound state $He\zeta$. If we denote
by $x$ the portion of dark matter composed of $UU\zeta$ (and
therefore $1-x$ the one made of $He\zeta$), by using
Eq.~({\ref{tbb}), we get the ratio $L/B$ as a function of $x$ as
\beq \frac{L}{B}=-3-5m_p \left( \frac{2x} {m_d
\sigma_{UU}}+ \frac{x}{m_d \sigma_{\zeta}}+\frac{1-x}{m_{\zeta} \sigma_{\zeta}} \right).
\label{dark} \eeq For our scenario to be realized, the quantity
inside the parenthesis of Eq.~({\ref{tbb}) must be negative, in
order to have abundance of $UU$ and not $\bar{U}\bar{U}$, while the quantity inside
the parenthesis of Eq.~(\ref{darko})~should be positive, in order to have more
$\zeta$ that $UU$. These two
constraints are satisfied by Eq.~(\ref{dark}). As we shall argue in
the next section, because of the strict CDMS constraints, $UU\zeta$
cannot be more than 4 to 6$\%$ of the dark matter density. The
overwhelming amount of dark matter in this case comes from
$He\zeta$. The result for the ratio $L/B$ is quite interesting. For
a large range of the parameters, the second term in the right hand
side of Eq.~(\ref{dark}) is much smaller than $-3$. This means that
our scenario makes a prediction for the ratio of leptons over the
baryons, independent of the specifics of the walking technicolor
model, i.e. the masses of the yet unknown hypothetical particles
$\zeta$ and $UU$, as well as other model dependent parameters like
the freeze out temperature of the sphalerons.

\begin{figure}[!tbp]
  \begin{center}

      \subfigure{\resizebox{!}{4.8cm}{\includegraphics{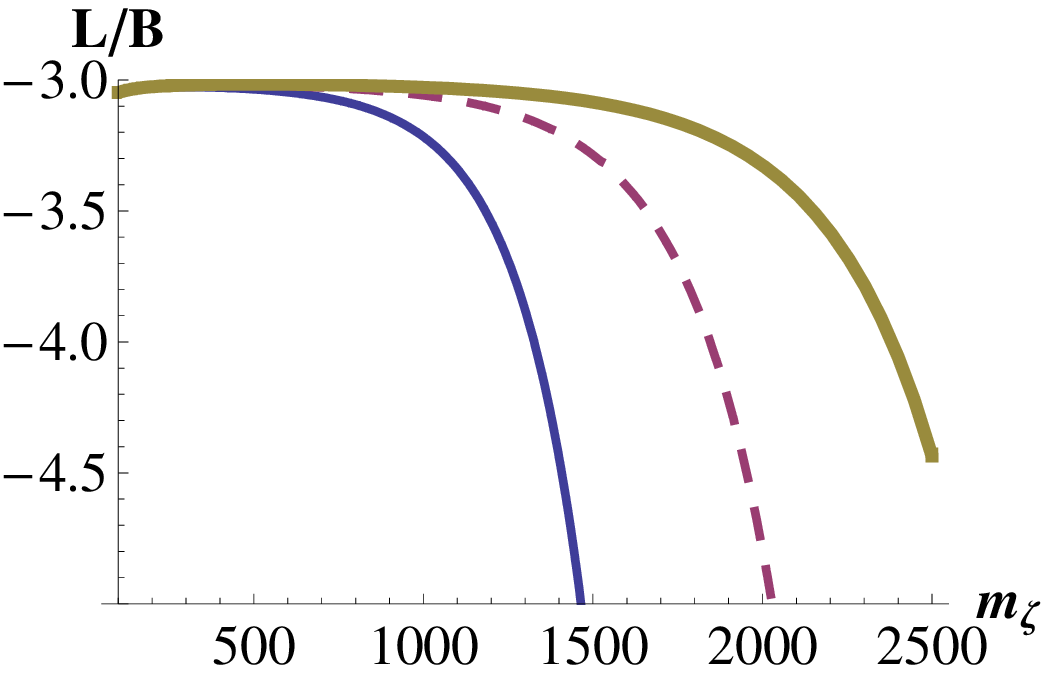}}} \quad
      \subfigure{\resizebox{!}{4.8cm}{\includegraphics{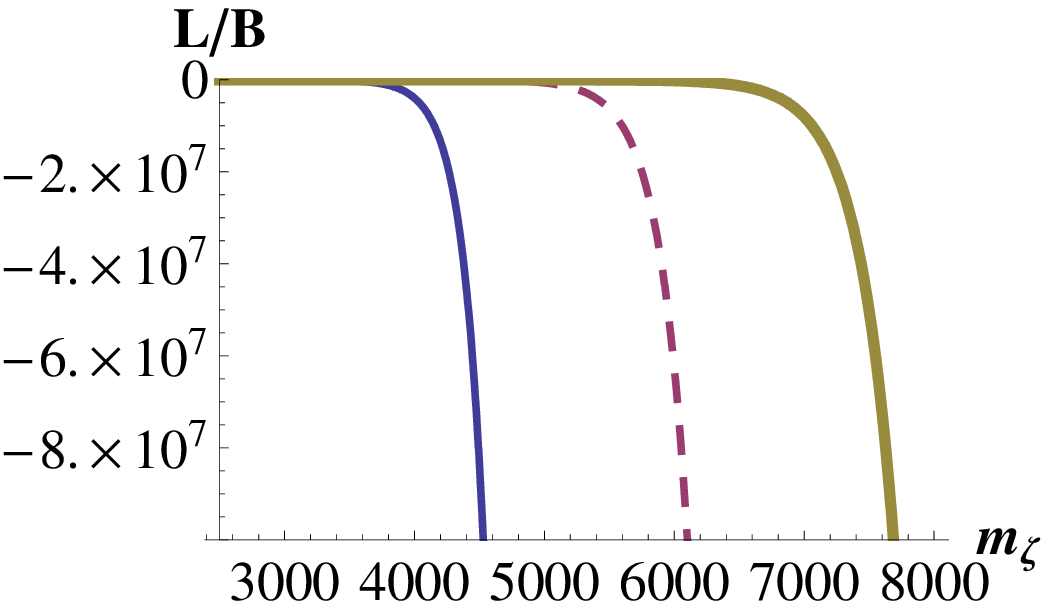}}} \\
      \subfigure{\resizebox{!}{4.8cm}{\includegraphics{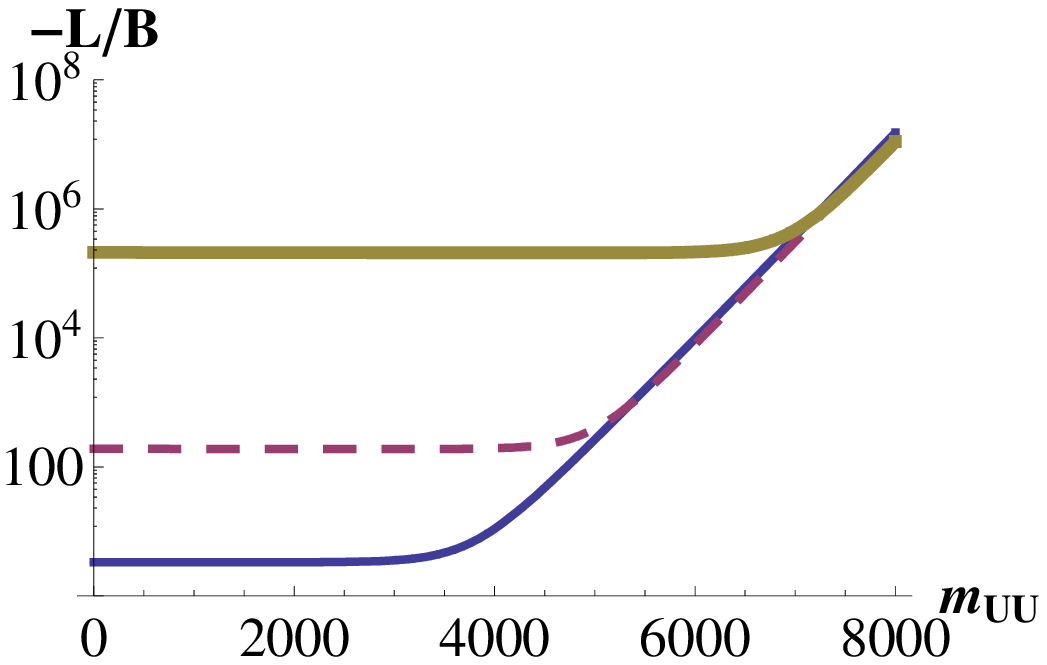}}} \quad
      \subfigure{\resizebox{!}{4.8cm}{\includegraphics{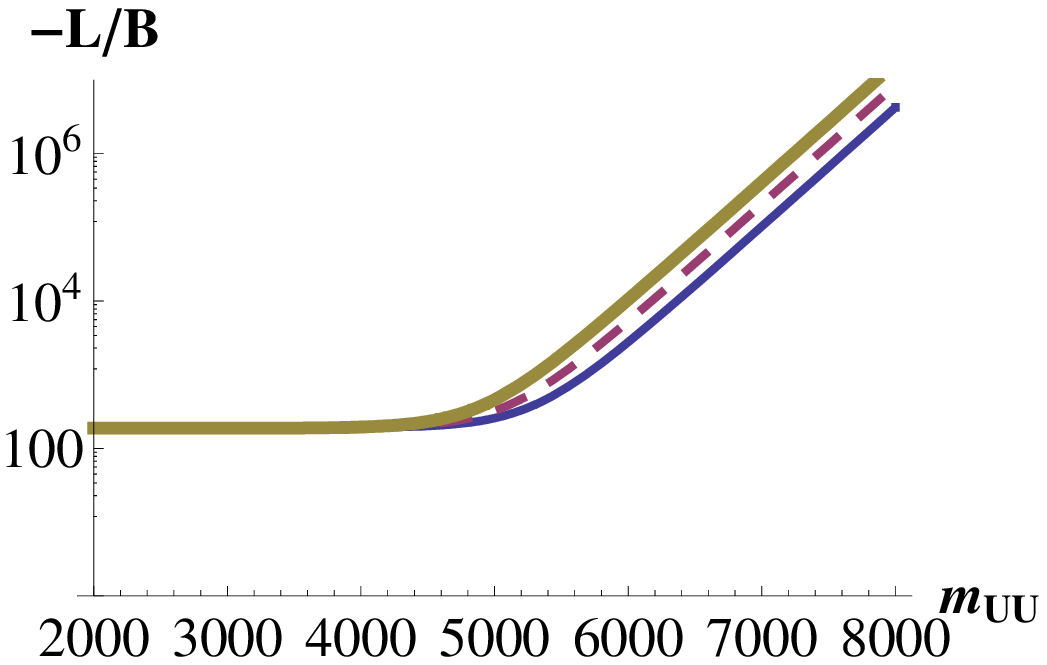}}}
    \caption{: The ratio $L/B$ derived from Eq.~({\ref{dark}). In
    the upper panels, $L/B$ is plotted as a function of $m_{\zeta}$, up to 2.5 TeV
    for the left panel and from 2.5 to 10 TeV for the right one. $x=0$ (there is
    no $UU\zeta$ dark
    matter density). The thin solid, dashed and thick solid lines
    correspond respectively to sphaleron freeze out temperatures
    of 150, 200, and 250 GeV. In the lower left panel, we plot the
    absolute value of $L/B$ in a logarithmic scale as a function
    of $m_{UU}$ if $UU\zeta$ makes up 3$\%$ of dark matter and
    sphaleron freeze out temperature is 250 GeV, for $m_{\zeta}=$
    2 TeV (thin solid line), 4 TeV (dashed line), and 6 TeV (thick
    solid line). In the lower right panel, we plot the same ratio as in
    the lower left panel having fixed $m_{\zeta}=4$ TeV, for three
    different values of $x$, $1\%$ (thin solid line), $2\%$ (dashed line), and
    $4\%$ (thick solid line).
    }}
    \label{fig3d}
    \end{center}
\end{figure}
In the upper left panel of Fig.~1, we show the predicted ratio of
$L/B$ as a function of the mass $m_{\zeta}$, if dark matter is
made only of bound states of $He\zeta$. In this case the mass of
$UU$ is irrelevant. We can see that the ratio $L/B$ is between
$-5$ to $-3$ independently of the values of $m_{UU}$ and
$m_{\zeta}$ as long as the latter does not exceed roughly 2 to 2.5
TeV. The predicted value for $L/B$ deviates rapidly from $-3$ as
soon as the mass of $\zeta$ becomes larger than roughly 2 to 2.5
TeV. In the upper right panel of Fig.~1, we show the $L/B$ ratio
(again with $x=0$) for $m_{\zeta}$ up to 10 TeV.
 $L/B$ increases (as an absolute value) exponentially for large
 values of $m_{\zeta}$. This is due to the fact that for large values
of $m_{\zeta}$, the statistical factor $\sigma_{\zeta}$
 becomes exponentially small. Therefore, the second
term in the right hand side of Eq.~(\ref{dark}) becomes much larger
than $-3$. Although one would expect $UU$ and $\zeta$ to have masses
of the order of the electroweak scale, due to lack of tools to
handle the non-perturbative dynamics of the technicolor model, at
the moment nothing forbids the masses to be even several TeV. $L/B$
becomes of the order of $\sim -10^8$ for $m_{\zeta}$ equal to
roughly 4.5, 6 and 7.5 TeV for sphaleron freeze out temperatures
150, 200 and 250 GeV respectively. Cosmological constraints forbid
larger negative values for $L/B$~\cite{Dolgov:2004jw} and therefore
this constrains $m_{\zeta}$.

Now we turn our interest to nonzero $x$, which means that there is
a small portion of WIMP type $UU\zeta$ dark matter. In the lower
left panel of Fig.~1, we plot the absolute value of $L/B$ in a
logarithmic scale as a function of $m_{UU}$ for the case where
$UU\zeta$ makes up 3$\%$ of the total dark matter density, for
three different values of $m_{\zeta}$ 2, 4, and 6 TeV. In the
lower right panel of Fig.~1, we plot the same ratio for a fixed
$m_{\zeta}=4$ TeV, and three different values of $x$, i.e. $1\%$,
$2\%$, and $4\%$.

\section{\label{Detection} Detection of composite states with techniparticles}
\subsection{Detection of ${\bf UU\zeta}$}
The neutral bound state between an electrically positively $+2$
charged $UU$ and a negatively $-2$ charged $\zeta$ has completely
different features as a dark matter candidate from $He\zeta$ and the
techni-O-helium candidates presented in~\cite{Khlopov:2007ic}. In
the case of $He\zeta$ (and in general for all techni-O-helium
candidates), because of the $He$ atom, the elastic cross section
with nuclei is very large (of the order of $10^{-25}~\cm^2$). If
such a particle exists, the large cross section with nuclei will
slow down the particle sufficiently in case it enters the atmosphere
of the Earth, that the recoil energy in the underground based detectors
like CDMS will be below the required
threshold~\cite{Khlopov:2007ic}. Only balloon, ground, or space based
detectors can possibly detect this particle. On the other hand,
$UU\zeta$ does not contain helium nucleus and has an elastic cross
section with nuclei much smaller than $He\zeta$. As we shall argue,
the elastic cross section is effectively the same as of a heavy
Dirac neutrino. This means that CDMS constraints should be taken
into consideration, since $UU\zeta$ behaves as a typical WIMP.

The elastic spin independent cross section of a neutral particle
scattering off nuclei targets is \beq
\sigma_0=\frac{G_F^2}{2\pi}\mu^2 Y^2 \bar{N}^2 F^2 \ , \eeq where
$G_F$ is the Fermi constant, $Y$ is the weak hypercharge of the WIMP
and $\mu$ is the reduced mass of the WIMP and the target nucleus.
$\bar{N} = N - (1-4sin^2\theta_w)Z$, where $N$ and $Z$ are the
number of neutrons and protons in the target nucleus and $\theta_w$
is the Weinberg angle. The parameter $F^2$ is a form factor for the
target nucleus. The cross section can be written as \beq \sigma_0 =
8.44 \times 10^{-3}\mu^2 Y^2 \bar{N}^2 F^2 {\rm pb} \ .
 \label{rate3}\eeq
In this case, our WIMP is not a single neutral particle, but it is
a bound state between two charged particles $UU^{++}$ and
$\zeta^{--}$. The Bohr radius of such a
bound state is of the order of $10^{-15}~\cm$ for typical
masses $m_{UU}$ and $m_{\zeta}$ of the order of TeV. For recoil
energies of the order of 10 keV, the $Z$ boson that mediates the
energy between the WIMP and the nucleus has a wavelength which is
of the same order of magnitude as the Bohr radius of $UU\zeta$.
This means that $Z$ interacts effectively with the whole $UU\zeta$
and not with the constituent particles $UU$ and $\zeta$. The
``effective'' hypercharge of $UU\zeta$ should be the sum of the
corresponding hypercharges of $UU$ and $\zeta$. The $UU$ has
hypercharge $+1$ since it belongs to the triplet of $UU$, $UD$,
and $DD$. The hypercharge of $\zeta$ is $-3/2$. Therefore the
``effective'' hypercharge of $UU\zeta$ is $Y=-1/2$. The $Ge$
detectors give the most strict constraints in CDMS so far. For a
Ge detector, $\bar{N}=38.59$.

For the form factor $F^2$, we use the Helm form factor \beq
F^2(q)=\left (\frac{3j_1(qR_1)}{qR_1}\right )^2e^{-q^2s^2}, \eeq
where $q=\sqrt{2M_nT}$ is the recoil momentum of the target
nucleus, $T$ is the recoil energy, $M_n$ is the mass of the target
nucleus, and $j_1(qR_1)$ is the spherical Bessel function. The
parameter $s=0.9$ fm and $R_1$ is defined through \beq
R_1=\sqrt{c^2+7\pi^2a^2/3-5s^2}, \eeq where $a=0.52$ fm and
$c\simeq 1.23A^{1/3}-0.6$ fm. For $Ge$ $A=73$. The number of
projected counts in CDMS is given by \beq {\rm counts} =
\int_{E_1}^{E_2} \frac{dR}{dT} dT \times \tau \ , \label{rate4}
\eeq where $\tau$ is the exposure of the detector measured in
kg.days, $E_1$ and $E_2$ are respectively the lower and upper
bounds for the recoil energy that the detectors have satisfying
efficiency. We take $E_1=10$ keV, $E_2=100$ keV, and $\tau=121$
kg.days. The differential rate with respect to the recoil energy
is \beq
  \label{rate1}
\frac{dR}{dT}=c_1\frac{R_0}{E_0 r}e^{-c_2T/E_0r},
 \eeq where $E_0$ is the kinetic energy
 of the WIMP, and $r=4mM_n/(m+M_n)^2$, $m$ being the mass of
the WIMP. The constants $c_1=0.751$, and $c_2=0.561$, are fitting
parameters that take into account the motion of the
Earth~\cite{Lewin:1995rx}. The parameter $R_0$ is \beq
R_0=\frac{503}{M_n m}\left(\frac{\sigma_0}{1{\rm
 pb}}\right)\left(\frac{\rho_{dm}}{0.4{\rm GeVc}^{-2}{\rm
 cm}^{-3}}\right)\left(\frac{\upsilon_0}{230{\rm kms}^{-1}}\right){\rm
 kg}^{-1}{\rm days}^{-1} \label{r0} .
\eeq The parameters $\rho_{dm}$ and $\upsilon_0$ are the local
dark matter density of the Earth and the thermal velocity of the
WIMP respectively. It is understood that since in our scenario
$UU\zeta$ makes up $x100\%$ of the total dark matter density,
$\rho_{dm}$ should always be multiplied by $x$. Here, we are going
to use $\upsilon_0=220~\text{km} / \text{s}$. We should also mention that the
$90\%$ confidence level for a confirmed event is 2.3 counts.
\begin{figure}[!tbp]
\begin{center}
\includegraphics[width=0.7\linewidth]{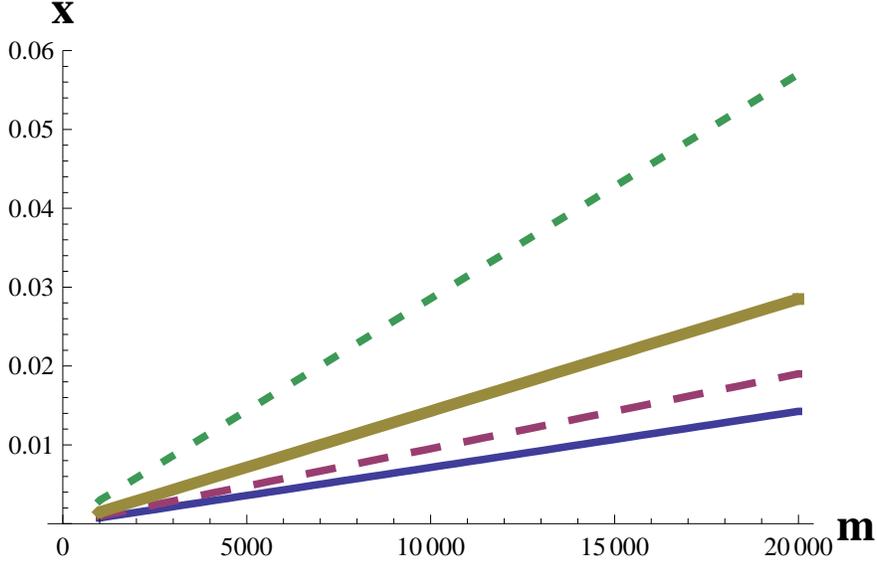}
\caption{The upper limit of the contribution of $UU\zeta$ to the
dark matter density as a function of its mass. The four lines
correspond to different local dark matter densities for the Earth,
namely  0.1 GeV$/\cm^3$ (dotted line), 0.2 GeV$/\cm^3$
(thick solid line), 0.3 GeV$/\cm^3$ (dashed line), and 0.4
GeV$/\cm^3$ (thin solid line). }
\end{center}
\end{figure}
So far, no confirmed counts have been found in
CDMS~\cite{Akerib:2005kh,Ahmed:2008eu}. As Fig.~2 shows, the CDMS
constraints restrict severely the percentage of $UU\zeta$ in the
dark matter. The area above the curves has been excluded by CDMS.
For a hypothetical $UU\zeta$ with a mass of 20 TeV, this particle
can only make up for 3 to 6$\%$ of the dark matter density (if the
local dark matter density of the Earth is between 0.1 and 0.3
GeV$/\cm^3$). A higher portion of $UU\zeta$ is allowed by
 the theory, but because of the spin independent cross
section with nuclei, it would have given a clean signal in CDMS.
This percentage of course depends strongly on the local dark
matter density and the mass of $UU\zeta$. The mass of $UU\zeta$ is
roughly speaking the sum of $m_{UU}$ and $m_{\zeta}$ and therefore
for a $UU\zeta$ mass of 20 TeV, it might correspond to particles
$UU$ and $\zeta$ with masses of 10 TeV. Under the circumstances,
$UU\zeta$ can be only a tiny fraction of dark matter and this
means that the rest of dark matter should be $He\zeta$. For
reasons that have been presented in~\cite{Khlopov:2007ic},
$He\zeta$ is not detectable in the detectors of CDMS. As we
already mentioned masses of $UU$ and/or $\zeta$ larger than 4.5 to
7.5 TeV (depending on the sphaleron freeze out temperature) give
an absolute value for the ratio $L/B$ larger than $10^8$ and this
might induce problems~\cite{Dolgov:2004jw}. The reason the curves
in Fig.~2 are almost straight lines has to do with the fact that
for large masses of $UU\zeta$, $R_0 \sim 1/m$, $E_0 \sim m$, and
$r \sim 1/m$. From
Eqs.~(\ref{rate4}),~({\ref{rate1}),~and~(\ref{r0}), we see that
the mass of $UU\zeta$ is proportional to the dark matter density
for a fixed number of counts.

Apart from the CDMS constraint regarding the mass of $UU$ and
$\zeta$, there is also a cosmological one. The existence of bound
states of $He\zeta$ reduces the abundance of free $He^4$ in the
Universe. The Standard Big Bang Nucleosynthesis (SBBN) predicts
roughly 25$\%$ abundance of $He$ among the baryons. The presence
of $He\zeta$ states requires that a portion of the free helium
will be captured by $\zeta$ in order to form $He\zeta$. If the
particle $\zeta$ is sufficiently light and if $He\zeta$ represents
a significant fraction of dark matter, a big amount of $He$ will
be captured, making the overall abundance of free helium much
smaller than the predicted (and observationally verified) 25$\%$.
The existence of systematic errors regarding the helium abundance
observed in the Universe is a legitimate possibility
\cite{book,Olive:2004kq}. For the derivation of our constraint, we
are going to assume that the helium abundance can be at most $\pm
2\%$ away from the SBBN predicted value. Upon making this
assumption, we can derive a lower bound for $m_{\zeta}$ as a
function of the percentage of $He\zeta$ in the overall dark matter
density. It is understood that the lighter $\zeta$ is, (given a
fixed mass density of $He\zeta$), the larger the number density of
the bound states of $He\zeta$ becomes and therefore the amount of
$He$ needed to form $He\zeta$. This means that the abundance of
free $He$ drops. The constraint comes exactly from the fact that
we do not allow a $He$ abundance decrease more that 2$\%$. The
densities of $He\zeta$ $n_{he\zeta}$~and $He$ $n_{He}$~are
respectively \beq \frac{\Omega_{He\zeta}}{\Omega_{B}}=5.5y
\Leftrightarrow n_{He\zeta}m_{He\zeta}=5.5y \Omega_{B},
\label{con1}\eeq \beq \frac{\Omega_{He}}{\Omega_{B}}=0.25
\Leftrightarrow n_{He}m_{He}=0.25 \Omega_{B}, \label{con2}\eeq
where $y=1-x$ represents the fraction of $He\zeta$ in the overall
dark matter density. The amount of $He$ captured by $\zeta$ is
equal to $n_{He\zeta}$ and according to the constraint it should
not be more than 2$\%$ of the overall baryon mass density.
Therefore, if we take the ratio of
Eqs.~(\ref{con1})~and~(\ref{con2}), we have \beq
\frac{n_{He\zeta}}{n_{He}}=\frac{5.5y}{0.25}\frac{m_{He}}{m_{He\zeta}}=\frac{2\%}{25\%}.
\eeq From this equation we can find the lowest $m_{\zeta}$ as a
function of $y$ that satisfies the constraint. The mass of
$He\zeta$ is roughly equal to the sum of $m_{He}$ and $m_{\zeta}$,
since the binding energy of $He\zeta$ is orders of magnitude
smaller than either one of them. From the CDMS constraint we
argued that $UU\zeta$ cannot account for more than 4 to 6$\%$ of
dark matter. This means that $He\zeta$ has to account for the rest
of the dark matter density. For an 100$\%$ component of $He\zeta$,
the lowest $m_{\zeta}=1022$~GeV. If $He\zeta$ is a component of
dark matter, this limit goes down accordingly. In any case, a
$\zeta$ heavier than roughly 1 TeV, no matter what is the amount
of $He\zeta$ in the dark matter density, change only slightly the
abundance of free $He$ and is consistent with the SBBN.

\subsection{Detection of techni-O-helium}
The constraint on WIMP-like $UU\zeta$ component of dark matter
leads to the scenario of techni-O-helium Universe, described
earlier in \cite{Khlopov:2007ic}. The composite nature of this
dominant fraction of techniparticle dark matter can lead to a
number of observable effects.

The nuclear interaction of techni-O-helium with cosmic rays gives
rise to ionization of this bound state in the interstellar gas and
to acceleration of free $\zeta^{--}$ in the Galaxy. Assuming a
universal mechanism of cosmic ray acceleration, the anomalous low
$Z/A$ component of $-2$ charged technileptons can be present in
cosmic rays and be within the reach for PAMELA and AMS02 cosmic ray
experiments.

It should be noted that techni-O-helium is not initially present
in significant amounts inside stars so that the injection of free
$\zeta$ from Supernova explosions might be suppressed, making the
regular mechanisms of cosmic ray acceleration ineffective for this
component. Then the $\zeta^{--}$ component may have such a low
momentum that it can be completely suppressed by Solar modulations
and can not penetrate heliosphere. On the other hand, at the stage
of red supergiant with size $\sim 10^{15} \cm$ during the period
$\sim 3 \cdot 10^{15} \s$ of this stage, up to $\sim 10^{-9}$ of
atoms of techni-O-helium per nucleon can be captured and give the
corresponding fraction in cosmic rays, accelerated by regular
mechanisms. 

Inelastic interaction of techni-O-helium with the matter in the
interstellar space can give rise to radiation in the range from
few keV to few  MeV. Though our first estimate shows that such a
radiation is below the cosmic nonthermal electromagnetic
background radiation observed in this range, special analysis of
this effect is of interest.

The evident consequence of the techni-O-helium dark matter is its
inevitable presence in the terrestrial matter. This is
because terrestrial matter appears opaque to $tOHe$ and stores all
its in-falling flux.

If the $tOHe$ diffusion in matter is determined by elastic
collisions, the in-falling $tOHe$ particles are effectively slowed
down after they fall down terrestrial surface. Then they drift, sinking down towards the center of
the Earth,
 with
velocity \beq V = \frac{g}{n \sigma v} \approx 8 S_2 A^{1/2} \cm/\s,
\label{dif}\eeq where $A \sim 30$ is the average atomic weight in
terrestrial surface matter, $n=2.4 \cdot 10^{24}/A$ is the number of
terrestrial atomic nuclei, $\sigma v$ is the rate of nuclear
collisions and $g=980~ \cm/\s^2$.

Near the Earth's surface, the techni-O-helium abundance is
determined by the equilibrium between the in-falling and
down-drifting fluxes. Such neutral $(^4He^{++}\zeta^{--})$ ``atoms"
may provide a catalysis of cold nuclear reactions in ordinary matter
(much more effectively than muon catalysis). This effect needs a
special and thorough investigation. On the other hand, $\zeta^{--}$
capture by nuclei heavier than helium~\cite{Cahn:1980ss}, can lead to production of
anomalous isotopes, but the arguments presented in
\cite{Khlopov:2007ic}, indicate that their abundance should be below
the experimental upper limits.

It should be noted that the nuclear cross section of the
techni-O-helium interaction with matter escapes the severe
constraints \cite{McGuire:2001qj} on strongly interacting dark
matter particles (SIMPs) \cite{McGuire:2001qj,Starkman} imposed by
the XQC experiment \cite{XQC}. Therefore, a special strategy of
techni-O-helium  search is needed, as it was proposed in
\cite{Belotsky:2006fa}.

In underground detectors, $tOHe$ ``atoms'' are slowed down to
thermal energies and give rise to energy transfer $\sim 2.5 \cdot
10^{-3} \eV A/S_2$, far below the threshold for direct dark matter
detection. It makes this form of dark matter insensitive to the
CDMS constraints. However, $tOHe$ induced nuclear transformation
can result in observable effects.

At a depth $L$ below the Earth's surface, the drift timescale is
$t_{dr} \sim L/V$, where $V \sim 40 S_2 \cm/\s$ is given by
Eq.~(\ref{dif}). It means that the change of the incoming flux,
caused by the motion of the Earth along its orbit, should lead at
the depth $L \sim 10^5 \cm$ to the corresponding change in the
equilibrium underground concentration of $tOHe$ on the timescale
$t_{dr} \approx 2.5 \cdot 10^3 S_2^{-1}\s$. Such rapid adjustment
of local fraction of $tOHe$ provides annual modulations of
inelastic processes inside the bodies of underground dark matter
detectors.

One can expect two kinds of inelastic processes in the matter,
composed of atoms with nuclei $(A,Z)$, having atomic number $A$ and charge $Z$
\beq (A,Z)+(He\zeta) \rightarrow (A+4,Z+2) +\zeta^{--},
\label{EHeAZ} \eeq
and
\beq (A,Z)+(He\zeta) \rightarrow [(A,Z)\zeta^{--}] + He.
\label{HeEAZ} \eeq
The first reaction is possible, if the masses of the initial and
final nuclei satisfy the energy condition \beq M(A,Z) + M(4,2) -
I_{o}> M(A+4,Z+2), \label{MEHeAZ} \eeq where $I_{o} = 1.6 \MeV$ is
the binding energy of techni-O-helium and $M(4,2)$ is the mass of
the $^4He$ nucleus. It is more effective for lighter nuclei, while
for heavier nuclei the condition (\ref{MEHeAZ}) is not valid and
reaction (\ref{HeEAZ}) should take place.

Both types of energy release processes are of the order of MeV,
which seems to have nothing to do with the signals in the DAMA
experiment. However, in the reaction (\ref{HeEAZ}) such energy is
rapidly carried away by the $He$ nucleus, while in the remaining
compound state of $[(A,Z)\zeta^{--}]$, the charge of the initial
$(A,Z)$ nucleus is reduced by 2 units and the corresponding
transformation of electronic orbits with possible emission of two
excessive electrons should take place. The energy difference between
the lowest lying $1s$ level of the initial nucleus with the charge
$Z$ and the respective levels of its compound system with
$\zeta^{--}$ is given by \beq \Delta E =Z^2\alpha^2m_e/2
-(Z-2)^2\alpha^2m_e/2 \approx Z\alpha^2m_e. \label{DEHeAZ} \eeq It
is interesting that the energy release in such a transition for
two $1s$ electrons in $^{53}I_{127}$ is about 2 keV, while for $^{81}Tl_{205}$
it is about 4 keV. Taking into account that the signal in the DAMA
experiment was detected with similar energy of ionization, this idea
deserves more detailed analysis, which might be useful for
interpretation of this experiment. Since the experimental cuts in
the CDMS experiment, exclude events of pure ionization, which are
not accompanied by phonon signal, if valid, the proposed mechanism
could explain the difference in the results of DAMA and CDMS.

\section{\label{Discussion} Discussion}
In this paper we explored the cosmological implications of a
 walking technicolor model with doubly charged technibaryons $UU^{++}$ and
technileptons $\zeta^{--}$. We studied a possibility for a
WIMP-like composite dark matter in the form of heavy ``atoms"
$[UU^{++}\zeta^{--}]$. To avoid overproduction of anomalous
isotopes (related to $UU^{++}$, which are not bound in these
atoms), the excess of $-2$ charged technileptons $\zeta^{--}$
should be larger than the excess of $UU^{++}$ generated in the
Universe. The residual doubly charged $\zeta^{--}$ bind with
$^4He$ in the
 techni-O-helium neutral states.

 In all the previous realizations of
composite dark matter scenarios, this excess was put by hand to
saturate the observed dark matter density. In our paradigm, the
abundance of techibaryons and technileptons is connected naturally
to the baryon relic density. Moreover, in a rather wide window of
techniparticle masses below few TeV, a robust prediction follows
for the ratio $L/B$ of lepton and baryon asymmetries. At further
increase of techniparticle mass, this ratio grows rapidly. It
provides an upper limit on the mass of techniparticles from the
condition that large negative value of $L/B$ does not lead to
overproduction of primordial $^4He$ in BBN.

Since techni-O-helium binds some fraction of $^4He$, an
interesting possibility appears that is at large values of $L/B$,
the excessive $^4He$ is hidden in the techni-O-helium. However,
due to the non-zero weak isospin charge of
$[UU^{++}\zeta^{--}]$, the presence of this dark matter component
should lead to observable effect in underground dark matter
detectors. The CDMS constraints reduce the allowed fraction of
this component to a few per cent, making techni-O-helium the
dominant form of composite dark matter in the considered scenario.
On that reason, a possibility to hide the excessive $^4He$ in the
techni-O-helium is elusive. On the contrary, even having taken
into account possible systematic errors in the determination of
primordial helium, to provide its abundance within the observed
limits, one should constraint the amount of helium bound with
$\zeta^{--}$. Since this amount is determined by the
techni-O-helium number density, the condition that techni-O-helium
saturates the observed dark matter density leads to a lower limit
for the mass of $\zeta^{--}$.

We come to the conclusion that in the minimal WTC model, contrary
to the case of the AC-model, WIMP-like component of composite
atom-like dark matter should be sparse, so that the formation of
large scale structure should follow a warmer than cold dark matter
scenario of the techni-O-helium Universe considered earlier.

In addition to the detailed description of a warmer than cold dark matter model, another challenging
 problem that is left for future work is the nuclear
transformations catalyzed by techni-O-helium. The question about
their consistency with observations remains open since special
nuclear physics analysis is needed to reveal what are the actual
techni-O-helium effects in BBN and in terrestrial matter.

The latter effects inside the body of underground dark matter detectors
can experience annual modulation and lead to ionization events
with a few keV energy release. It can make techni-O-helium (as well as
any other form of O-helium) an interesting candidate, which might explain the
difference between the positive result of DAMA/NaI (DAMA/Libra) and
negative results of other experiments on direct dark matter search.

The destruction of techni-O-helium by cosmic rays in the Galaxy
releases free charged technileptons, which can be accelerated and
contribute to the flux of cosmic rays. In this context, the search
for techniparticles at accelerators and in cosmic rays acquires
the meaning of a crucial test for the existence of the basic
components of the composite dark matter. At accelerators,
techniparticles would look like stable doubly charged heavy
leptons, while in cosmic rays, they represent a heavy $-2$ charge
component with anomalously low ratio of electric charge to mass.
If it has the same energy spectrum as ordinary cosmic rays, it can
be observed in the PAMELA experiment.

To conclude, the minimal walking technicolor cosmology can give a robust
cosmological scenario of composite dark matter, giving rise to a set of
exciting observable effects.

\section*{Acknowledgements}
We express our gratitude to P.~Belli, K.M.~Belotsky, J.~Filippini, A.G.~Mayorov,
P.~Picozza, V.A.~Rubakov and E.Yu.~Soldatov for important comments and for useful
discussions. The work of CK was supported by the Marie Curie
Fellowship under contract MEIF-CT-2006-039211.


\begin{thebibliography}{199}

\bibitem{Khlopov:2007ic}
  M.~Y.~Khlopov and C.~Kouvaris,
  Phys.\ Rev.\  D {\bf 77}, 065002 (2008)
  [arXiv:0710.2189 [astro-ph]].

\bibitem{Sannino:2004qp}
F.~Sannino and K.~Tuominen,
 Phys.\ Rev.\ D {\bf 71}, 051901 (2005);
arXiv:hep-ph/0405209.

\bibitem{Hong:2004td}
  D.~K.~Hong, S.~D.~H.~Hsu and F.~Sannino,
    Phys.\ Lett.\ B {\bf 597}, 89 (2004);
  arXiv:hep-ph/0406200.


\bibitem{Dietrich:2005jn}
  D.~D.~Dietrich, F.~Sannino and K.~Tuominen,
  Phys.\ Rev.\ D {\bf 72}, 055001 (2005);
  arXiv:hep-ph/0505059.

\bibitem{Dietrich:2006cm}
  D.~D.~Dietrich and F.~Sannino,
  Phys.\ Rev.\  D {\bf 75}, 085018 (2007)
  [arXiv:hep-ph/0611341].

\bibitem{Gudnason:2006mk}
  S.~B.~Gudnason, T.~A.~Ryttov and F.~Sannino,
  Phys.\ Rev.\  D {\bf 76}, 015005 (2007)
  [arXiv:hep-ph/0612230].


\bibitem{Gudnason:2006ug}
  S.~B.~Gudnason, C.~Kouvaris and F.~Sannino,
  Phys.\ Rev.\  D {\bf 73}, 115003 (2006);
  arXiv:hep-ph/0603014.

\bibitem{Foadi:2007ue}
  R.~Foadi, M.~T.~Frandsen, T.~A.~Ryttov and F.~Sannino,
  Phys.\ Rev.\  D {\bf 76}, 055005 (2007)
  [arXiv:0706.1696 [hep-ph]].

\bibitem{Dietrich:2008ni}
  D.~D.~Dietrich and C.~Kouvaris,
  arXiv:0805.1503 [hep-ph].

\bibitem{Catterall:2007yx}
  S.~Catterall and F.~Sannino,
  Phys.\ Rev.\  D {\bf 76}, 034504 (2007)
  [arXiv:0705.1664 [hep-lat]].

\bibitem{DelDebbio:2008wb}
  L.~Del Debbio, M.~T.~Frandsen, H.~Panagopoulos and F.~Sannino,
  arXiv:0802.0891 [hep-lat].

\bibitem{DelDebbio:2008zf}
  L.~Del Debbio, A.~Patella and C.~Pica,
  arXiv:0805.2058 [hep-lat].

\bibitem{Gudnason:2006yj}
  S.~B.~Gudnason, C.~Kouvaris and F.~Sannino,
  Phys.\ Rev.\  D {\bf 74}, 095008 (2006);
  arXiv:hep-ph/0608055.

\bibitem{Kouvaris:2007iq}
  C.~Kouvaris,
  Phys.\ Rev.\  D {\bf 76}, 015011 (2007)
  [arXiv:hep-ph/0703266].

\bibitem{Kainulainen:2006wq}
  K.~Kainulainen, K.~Tuominen and J.~Virkajarvi,
  Phys.\ Rev.\  D {\bf 75}, 085003 (2007)
  [arXiv:hep-ph/0612247].

\bibitem{Glashow:2005jy}
  S.~L.~Glashow,
  arXiv:hep-ph/0504287.

\bibitem{Fargion:2005xz}
  D.~Fargion and M.~Khlopov,
  arXiv:hep-ph/0507087.

\bibitem{Fargion:2005ep}
  D.~Fargion, M.~Khlopov and C.~A.~Stephan,
  Class.\ Quant.\ Grav.\  {\bf 23}, 7305 (2006)
  [arXiv:astro-ph/0511789].

\bibitem{Khlopov:2006uv}
  M.~Y.~Khlopov and C.~A.~Stephan,
  arXiv:astro-ph/0603187.

\bibitem{Connes:1994yd}
  A.~Connes,
  ``Noncommutative geometry", Academic Press, London and San Diego, 1994.

\bibitem{Stephan:2005uj}
  C.~A.~Stephan,
  J.\ Phys.\ A  {\bf 39}, 9657 (2006)
  [arXiv:hep-th/0509213].

\bibitem{Khlopov:2005ew}
  M.~Y.~Khlopov,
  Pisma Zh.\ Eksp.\ Teor.\ Fiz.\  {\bf 83}, 3 (2006)
  [JETP Lett.\  {\bf 83}, 1 (2006)]
  [arXiv:astro-ph/0511796].

\bibitem{Belotsky:2006fd}
  K.~Belotsky, M.~Khlopov and K.~Shibaev,
  arXiv:astro-ph/0602261.

\bibitem{Belotsky:2006pp}
  K.~M.~Belotsky, M.~Y.~Khlopov and K.~I.~Shibaev,
  Grav.\ Cosmol.\  {\bf 12}, 93 (2006)
  [arXiv:astro-ph/0604518].

\bibitem{4Q}
K.~M.~Belotsky, M.~Y.~Khlopov and K.~I.~Shibaev,
Stable quarks of the 4th family?
invited contribution to the book "The Physics of
 Quarks: New Research." to be published by NOVA, 2008.

\bibitem{Belotsky:2005ui}
  K.~M.~Belotsky, M.~Y.~Khlopov, K.~I.~Shibaev, D.~Fargion, R.~V.~Konoplich and M.~G.~Ryskin,
  Grav.\ Cosmol.\  {\bf 11} (2005) 3.

\bibitem{Khlopov:2006dk}
  M.~Y.~Khlopov,
  arXiv:astro-ph/0607048.

\bibitem{Khlopov:2007zza}
  M.~Y.~Khlopov,
{\it Prepared for 10th Workshop on What Comes Beyond the Standard Model, Bled, Slovenia, 17-27 Jul 2007}

\bibitem{Khlopov:2008rp}
  M.~Y.~Khlopov,
  arXiv:0801.0167 [astro-ph].

\bibitem{Khlopov:2008rq}
  M.~Y.~Khlopov,
  arXiv:0801.0169 [astro-ph].

\bibitem{exp3}
P.~Mueller, Phys.Rev.Lett. {\bf 92}, 22501 {2004};
arXiv:nucl-ex/0302025.
%
\bibitem{Akerib:2005kh}
  D.~S.~Akerib {\it et al.}  [CDMS Collaboration],
  Phys.\ Rev.\ Lett.\  {\bf 96}, 011302 (2006)
  [arXiv:astro-ph/0509259].

\bibitem{Ahmed:2008eu}
  Z.~Ahmed {\it et al.}  [CDMS Collaboration],
  arXiv:0802.3530 [astro-ph].

\bibitem{Bernabei:2003za}
  R.~Bernabei {\it et al.},
  Riv.\ Nuovo Cim.\  {\bf 26N1}, 1 (2003)
  [arXiv:astro-ph/0307403].

\bibitem{Bernabei:2008yi}
  R.~Bernabei {\it et al.}  [DAMA Collaboration],
  arXiv:0804.2741 [astro-ph].

\bibitem{Dolgov:2004jw}
  A.~D.~Dolgov and F.~Takahashi,
  Nucl.\ Phys.\  B {\bf 688} (2004) 189
  [arXiv:hep-ph/0402066].


\bibitem{Lewin:1995rx}
  J.~D.~Lewin and P.~F.~Smith,
  Astropart.\ Phys.\  {\bf 6}, 87 (1996).

\bibitem{book}
  M.~Y.~Khlopov,
  ``Cosmoparticle physics,''
{\it  Singapore: World Scientific (1999) 577 p}

\bibitem{Olive:2004kq}
  K.~A.~Olive and E.~D.~Skillman,
  Astrophys.\ J.\  {\bf 617}, 29 (2004)
  [arXiv:astro-ph/0405588].

\bibitem{Cahn:1980ss}
  R.~N.~Cahn and S.~L.~Glashow,
  Science {\bf 213}, 607 (1981).

\bibitem{McGuire:2001qj}
B.\,D. Wandelt et al.,
  ``Self-interacting dark matter,''
  arXiv:astro-ph/0006344;
P.\,C. McGuire and P.\,J. Steinhardt,
  ``Cracking open the window for strongly interacting massive particles as  the
  halo dark matter,''
  arXiv:astro-ph/0105567;
 G. Zaharijas and G.\,R. Farrar,
  Phys.~Rev. {\bf D72}, 083502 (2005);
  arXiv:astro-ph/0406531.

\bibitem{Starkman}
  C.\,B. Dover, T.\,K. Gaisser and G. Steigman,
  Phys.~Rev.~Lett. {\bf 42}, 1117 (1979);
  S. Wolfram,
  Phys.~Lett.  {\bf B82}, 65 (1979);
G.\,D. Starkman et al., Phys.~Rev. {\bf D41}, 3594 (1990);
 D.~Javorsek et al.,
  Phys.~Rev.~Lett. {\bf 87}, 231804 (2001);
S. Mitra,
  Phys.~Rev. {\bf D70}, 103517 (2004);
  arXiv:astro-ph/0408341.
Mack, G.D.; Beacom, J.F.; Bertone, G.
  Phys Rev 2007, vol D76, 043523;
  arXiv:0705.4298 [astro-ph].

\bibitem{XQC}
D. McCammon  et al., Nucl.~Instr.~Meth. {\bf A370}, 266 (1996); D.
McCammon et al.,
  Astrophys.~J. {\bf 576}, 188 (2002);
  arXiv:astro-ph/0205012.

\bibitem{Belotsky:2006fa}
  K.~Belotsky, Yu.~Bunkov, H.~Godfrin, M.~Khlopov and R.~Konoplich,
  ``He-3 experimentum crucis for dark matter puzzles,''
  arXiv:astro-ph/0606350.


\end{thebibliography}
\end{document}